# Characterization of Prototype BTeV Silicon Pixel Sensors Before and After Irradiation

Maria R. Coluccia[1], J. A. Appel[1], G. Chiodini[1], D. C. Christian[1], and S. Kwan[1]

*Abstract*-- We report on measurements performed on silicon pixel sensor prototypes exposed to a 200 MeV proton beam at the Indiana University Cyclotron Facility. The sensors are of $n^+/n/p^+$ type with multi-guard ring structures and p-stop electrode isolation on the $n^+$-side. Electrical characterization of the devices was performed before and after irradiation up to a proton fluence of $4 \times 10^{14}$ p/cm$^2$. We tested pixel sensors fabricated from normal and oxygen-enriched silicon wafers and with two different p-stop isolation layouts: common p-stop and individual p-stop.

## I. INTRODUCTION

BTeV is an experiment expected to run in the new Tevatron C0 interaction region (IR) at Fermilab. It is designed to perform precision studies of *b* and *c* quark decays, with particular emphasis on mixing, CP violation, and rare and forbidden decays [1]. An important feature of BTeV is that a detached vertex trigger algorithm is implemented in the first level trigger [2]. Consequently, the vertex detector must have superior pattern-recognition power, small track extrapolation errors, and good performance even after high radiation dose. Silicon pixel sensors were chosen because they provide very accurate space point information and have intrinsically low noise and high radiation hardness.

The baseline BTeV silicon pixel detector [1] has rectangular 50 µm × 400 µm pixel elements. It has doublets of planes distributed along the IR separated by 4.25 cm. Half-planes are mounted above and below of the beam, and are arranged so that a small square hole of ±6 mm × ±6 mm is left for the beam to pass through. At such small distance from the colliding beams, the pixel detectors will be exposed to a significant level of irradiation. At the full luminosity at which we plan to operate, the innermost pixel detector will receive a fluence of $1 \times 10^{14}$ minimum ionization particles/cm$^2$/year (~$0.5 \times 10^{14}$ 1-MeV neutron equivalent/cm$^2$/year). This will lead to radiation damage to both the surface and the bulk of the silicon pixel sensors.

The bulk damage is mainly due to the non-ionizing energy loss (NIEL), which, through the displacement of the atoms in the crystal lattice, creates new energy levels, effectively acting as acceptors. Therefore, the effective doping concentration will change with irradiation. This will eventually lead to the inversion of the conduction type of the bulk material (type-inversion), increases in leakage current and depletion voltage, changes in capacitance and resistivity, and charge collection losses [3]. These are problems that need to be addressed by all the next generation hadron collider experiments. As a result, there is a worldwide effort to address these technical challenges. Solutions include the design of multiple guard ring structures to avoid avalanche breakdown along the edge [4]-[5], low resistivity silicon substrates to delay type inversion [6], and oxygenated silicon wafers to reduce the effects of radiation-induced formation of defects in the silicon lattice [7].

In order to increase the useful operating time of the silicon sensors, running with partial depletion has to be considered. Such operation might be necessary if the full depletion voltage becomes excessively large after the substrate type inversion. For this reason, the BTeV pixel sensors have $n^+/n/p^+$ configuration with pixels on the $n^+$ side. After type inversion, the depleted region grows from the $n^+$ side of the junction and the sensor can operate partially depleted. However, for $n^+/n$ devices, it is necessary to have an electrical isolation between neighboring cells to maintain high resistance in the presence of the electron accumulation layer at the silicon/silicon-dioxide interface. There are two isolation technologies: the p-stop technique [8], in which a high dose ($> 10^{13}$/cm$^2$) p-type implant surrounds each $n^+$-type region, and the p-spray technique [9], in which there is an application of medium (~$3.0 \times 10^{12}$/cm$^2$) dose p-type implant to the whole n-side.

In this first phase of our studies, we tested only prototype sensors with p-stop electrode isolation. The ideal design of the p-stop needs to be studied because it has a significant impact on the minimum pixel pitch, the noise, charge collection, capacitance, and breakdown voltage [10].

## II. DEVICE STRUCTURES

We have received detectors from SINTEF Electronics and Cybernetics (Oslo, Norway). The base material is low resistivity (1.0-1.5 kΩ cm) <100> silicon, 270µm thick. Some of the wafers have been oxygenated. The oxygenation process is done in a N$_2$ environment for 72 hours at 1150 °C.

Manuscript received November 23, 2001. Work supported by U.S. Department of Energy under contract no. DE-AC02-76CH0300.
[1]Fermi National Accelerator Laboratory, P.O. Box 500 Batavia, IL 60510 USA.

The tested devices consist of silicon pixel sensors having two different layouts of p-stop electrode isolation: individual and common p-stops. For the individual p-stops there is a p-implant ring (atoll) around each pixel. For the common p-stop, there is a continuous p-implant along pixel columns and rows (see Fig. 1 for details).

We tested two pixel array sizes with single cell dimension 50 μm × 400 μm. The first array (called "test-sized sensor") contains 12 × 92 cells. The second array (called "FPIX1-sized sensor") contains 18 × 160 cells and it is designed to be read out by a single front-end chip FPIX1 [11]. These two arrays are characterized also by variations in implant widths. In Fig. 1 we show a drawing of the two different p-stop layouts and in Table I we list the values of the implant widths and the separations between the implants for the two pixel arrays.

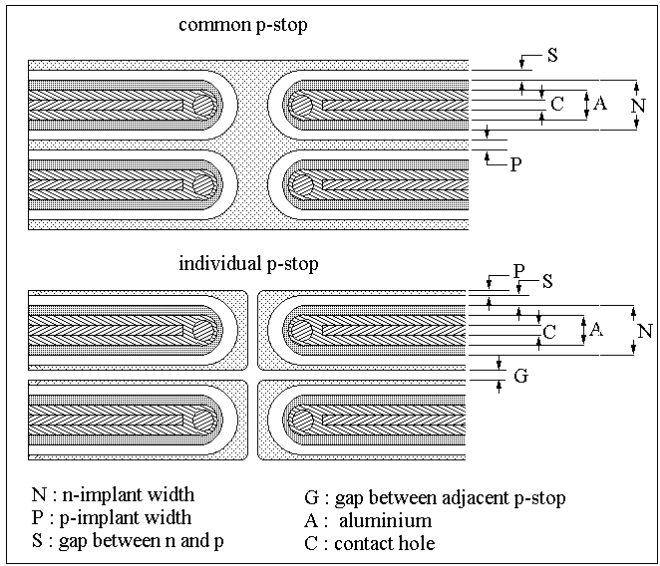

Fig. 1. Common p-stop and individual p-stop pixel sensor layouts.

TABLE I
SUMMARY OF THE IMPLANT WIDTHS FOR THE PIXEL ARRAYS

| sensor | p-stop design | N n-implant width (μm) | P p-implant width (μm) | S gap between n and p (μm) | G gap between adjacent p-stop (μm) |
|---|---|---|---|---|---|
| test-sized | common | 23 | 9 | 9 | \ |
| FPIX1-sized | common | 31 | 3 | 8 | \ |
| test-sized | individual | 23 | 5 | 6 | 5 |
| FPIX1-sized | individual | 21 | 5 | 8 | 3 |

Comparisons among different guard ring structures are also reported. We tested three different p-side guard ring structures. For the test-sized sensors, we have two guard ring structures having 10 and 18 rings respectively. For the FPIX1-sized sensors we have 11 guard rings. In the case of 10 guard rings, every ring has 15 μm $p^+$ implantation, 23 μm of metalization (that overlaps the $p^+$ implant by 4 μm on both sides) and 11 μm of passivation opening [12]-[13]. There is a large $n^+$ region between the last guard ring and the scribe line. Going outwards from bias ring toward the $n^+$ region, the gaps among adjacent rings increase from 15 μm to 30 μm. For the structures with 11 and 18 rings, we adopted the same design as described in [4] and implemented in the ATLAS prototype I pixel sensor design [14]. Each ring has a p-implant 10 μm wide and the pitch increases from 20 μm for the innermost ring to 50 μm near the edge of the detector. In addition there is a metal field plate that overhangs the p-implant and extends inwards by half the gap width towards the active area [15].

### III. EXPERIMENTAL PROCEDURES

Electrical characterization of the devices was performed with standard techniques (I-V, $V_{g\text{-ring}}$-V and C-V curves) before and after irradiation. We used a Keithley 237 as power supply and current monitor, and both QTech 7600 and HP4274A LRC meters for the C-V measurements.

All the measurements were performed using a probe station placed in a dark box in a clean room. Continuous monitoring of temperature and humidity were performed, and all the measurements reported were done at 0 % relative humidity, achieved by flowing dry nitrogen in the dark box. In order to investigate the stability of the electrical characteristics, several measurements were performed in various humidity conditions (ranging from 0 % to 40 %), but no significant difference was detected.

The measurements were performed with the p-side (sensor back-plane) negatively biased through one probe and the n-side grounded through the chuck. We measured the leakage current and the capacitance for the whole sensor without considering the contribution from the guard rings. We performed some measurements before irradiation biasing the innermost guard ring together with the p-side but we found that the contribution was negligible.

Twelve wafers (three oxygenated), each containing six test-sized sensors and seven FPIX1-sized sensors, were characterized before and after dicing. Several of the single devices were characterized before and after irradiation at the Indiana University Cyclotron Facility (IUCF). We report results obtained from both oxygenated and non-oxygenated wafers and individual devices.

The proton irradiation tests took place with a 200 MeV proton beam. The displacement damage cross section for 200 MeV protons (90.5 MeVmb) [16] is almost exactly the same as the value conventionally assigned to 1 MeV neutrons (95 MeVmb) [17] so we quote our results as a function of proton fluence rather than equivalent 1 MeV neutron fluence. The beam profile was measured by exposing a sensitive film. The beam spot, defined by the circular area where the flux is within 90 % of the central value, had a diameter of 1.5 cm, comfortably larger than the sensor size (the FPIX1-sized sensors is ~1 cm x 1 cm). Before the exposure, the absolute fluence was measured with a Faraday cup; during the exposure the relative fluence was determined with a Secondary Electron Emission Monitor.

We used a PC board with a big opening in the middle (4 inch × 4 inch) where we placed the sensors with simple cardboard supports. The irradiation was done in air at room temperature and took no more than six hours. The exposures with multiple boards were done placing the boards about 2 cm behind each other and with the pixel side facing the beam. Mechanically, the boards were kept in position by an open aluminum frame. A maximum of six boards were exposed each time, and therefore the beam energy degradation was negligible. After irradiation, the tested devices were kept at minus 15 °C in order to slow down the reverse annealing process.

The measurements after irradiation were performed in a condition in which the plateau of the beneficial annealing has not been reached. We are interested in investigating the behavior of the sensors in an environment that is as close as possible to the real experiment. The operational temperature of the vertex detector in BTeV will be between –5 °C and –10 °C and therefore the pixel sensors will not profit from beneficial annealing. For this reason, we decided to store the irradiated sensors at low temperature just after irradiation. The measurements at room temperature took no more than a few hours.

Typically, measurements were made thirty days after irradiation.

## IV. RESULTS AND DISCUSSION

### A. Basic Sensor Performance Before Irradiation

We present first a comparison between individual and common p-stop pixel sensors.

Fig. 2 shows the typical I-V curves for test-sized sensors from a non-oxygenated wafer before dicing. We made measurements on five different wafers (20 sensors in total) and, apart from a few sensors which show higher leakage current and relatively lower breakdown voltage, we found the same results for both types of pixel isolation layout. One would expect the individual p-stop layout to show lower breakdown voltage due to the presence of an electrical field gradient along the "atoll" that is not presents in the common p-stop layout. However, from our measurement of the leakage current and breakdown voltage, no significant differences were detected between the two isolation layouts. We also do not see any difference between the two guard ring structures.

The breakdown voltage distribution for the test-sized sensors of these wafers has a median value around 700 V and this, together with the fact that the current is very small (~10 nA/cm$^2$ after depletion), shows the good performance of these sensors. Fig. 3 shows the breakdown voltage of different test-sized common and individual p-stop sensors from various oxygenated (right two groups) and non-oxygenated wafers (left five groups). Only a few sensors have poor performance. The yield for this SINTEF wafer series is very high.

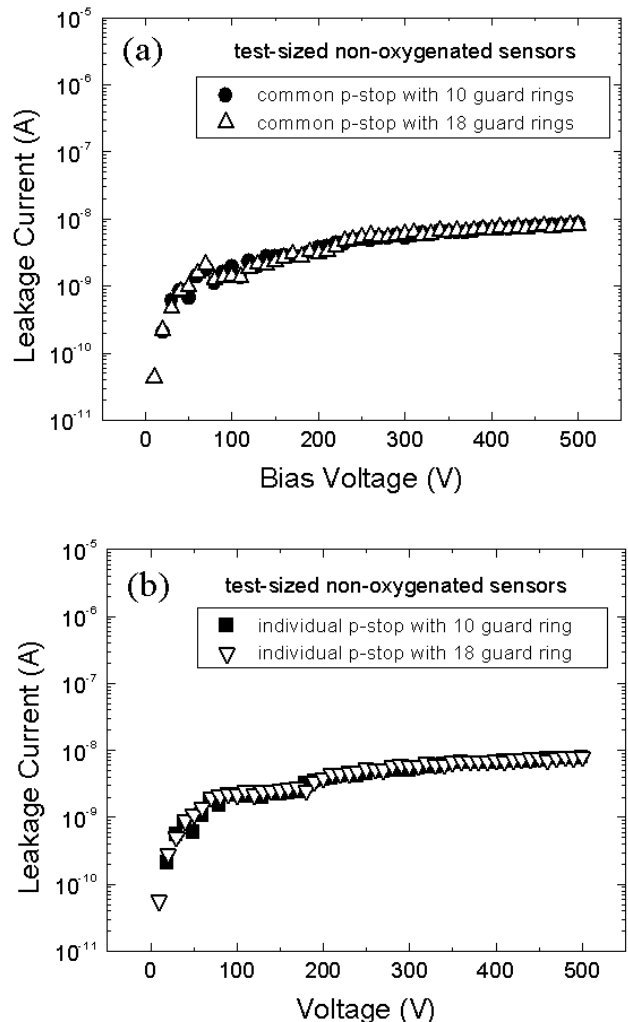

Fig. 2. I-V characteristics for un-irradiated test-sized pixel sensors from one non-oxygenated wafer: a) common p-stop pixel sensors, b) individual p-stop pixel sensors.

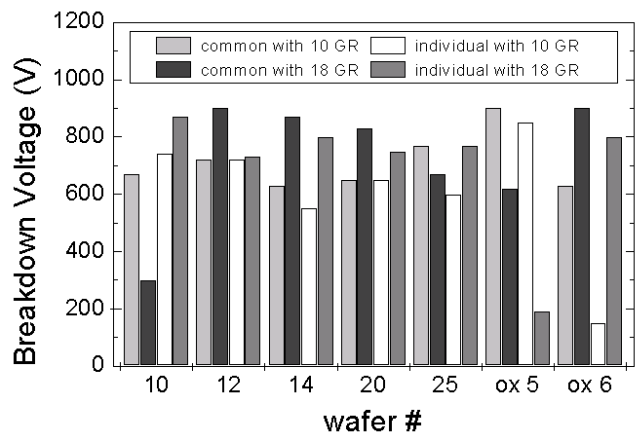

Fig. 3. Breakdown voltage for common and individual p-stop test-sized pixel sensors from oxygenated and normal SINTEF wafers. We present the result for the two different guard ring structures.

We also measured the CV curve for each sensor in order to determine the full depletion voltage (and therefore the operating voltage). The depletion voltage is normally defined as the bias voltage required so that the region depleted of free carriers reaches through the whole of the semiconductor bulk and is extracted from the C-V curves as the intersection point of two fitted straight lines in the logC-logV plot. Typically the measurements are performed at a frequency of 50 kHz and at room temperature (~24 °C) using the QTech 7600 LRC meter. However, some of the measurements were performed using the HP4274A for which only few values of frequency are provided and we chose to perform the measurement at a frequency of 40 kHz. The typical depletion voltage before irradiation is around 210 V with a small spread.

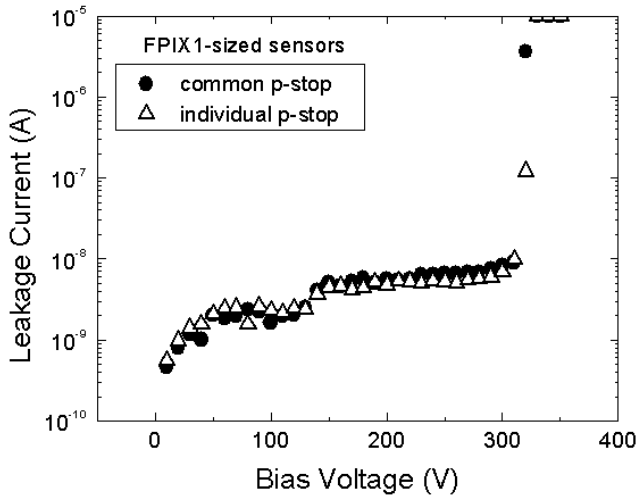

Fig. 4. I-V characteristics for FPIX1-sized sensors from a non-oxygenated wafer.

Fig. 4 shows the typical I-V curves for two FPIX1-sized sensors (one common p-stop and one individual p-stop) from one non-oxygenated wafer. We tested 15 FPIX1-sized common p-stop sensors and 20 FPIX1-sized individual p-stop sensors. Apart from a few sensors with slightly poorer performances (early onset of breakdown) most sensors tested show the same I-V characteristic.

Comparing Fig. 2 and Fig. 4, one can see that the breakdown voltage for the FPIX1-sized sensors is lower (typically just above 300V) than the test-sized sensors (700 V). This is likely due to the fact that the single cell in these sensors is characterized by different p implant widths. In fact, from Table I, we can see that for the FPIX1-sized individual p-stop sensors the separation between two adjacent p-stop rings is 3 µm instead of 5 µm for the individual p-stop test-sized sensors. For the FPIX1-sized common p-stop sensors the p-implant width is also 3 µm instead of 9 µm in the common p-stop test-sized sensors. The narrow line width does not conform to the design rules as specified by the vendor.

Some of the wafers were then diced using a diamond-dicing saw (Disco-DAD320). After dicing, we observed an increase in the leakage current and earlier onset of breakdown with respect to the un-diced sensors. By carefully cleaning the surface with acetone and deionized water, we can restore the performance that we had before dicing. This is due to the fact that the dicing process introduces impurities and silicon debris on the sensor surface and on the edges. These can be eliminated with proper cleaning.

We have also studied the I-V curve of the FPIX1-sized sensors after bump bonding to a readout chip. We observed similar I-V and breakdown voltage as for the bare sensors.

### B. Performance After Irradiation

We irradiated six test-sized and four FPIX1-sized sensors. Here we present not only a detailed study of the sensor behavior after irradiation, but also a comparison between individual and common p-stop pixel isolation, among various guard rings structures, and between oxygenated and non-oxygenated sensors. Table II summarizes the features of the sensors that we irradiated and shows also the fluence received by each of them. The results for the oxygenated sensors will be discussed in the next section.

TABLE II
SUMMARY OF THE DEVICES USED FOR THE IRRADIATION TEST

| sensor | p-stop design | oxygenation | guard ring # | breakdown voltage (V) after irrad. | fluence (p cm$^{-2}$) |
|---|---|---|---|---|---|
| test-sized | common | no | 18 | >500 | 8 x 10$^{13}$ |
| test-sized | individual | no | 18 | >500 | 8 x 10$^{13}$ |
| FPIX1-sized | common | no | 11 | 250 | 2 x 10$^{14}$ |
| FPIX1-sized | individual | no | 11 | 180 | 2 x 10$^{14}$ |
| FPIX1-sized | common | yes | 11 | 220 | 2 x 10$^{14}$ |
| FPIX1-sized | individual | yes | 11 | 220 | 2 x 10$^{14}$ |
| test-sized | common | no | 10 | >500 | 4 x 10$^{14}$ |
| test-sized | individual | no | 10 | >500 | 4 x 10$^{14}$ |
| test-sized | common | yes | 18 | >500 | 4 x 10$^{14}$ |
| test-sized | individual | yes | 10 | >500 | 4 x 10$^{14}$ |

Fig.5 shows the leakage current measurements before and after irradiation for the common p-stop sensors. We found the same results also for the individual p-stop sensors that we irradiated. After irradiation the leakage current increases by several orders of magnitude, and, as expected, shows a nearly linear dependence on fluence. In fact, the increase of the leakage current $\Delta I$ (i.e. the difference between the currents measured after and before irradiation) shows a linear dependence on the fluence: $\Delta I = \alpha \Phi V$ where $\alpha$ is the damage constant, $\Phi$ is the fluence, and V is the volume. Fig. 6 shows the fluence dependence of the increase in leakage current normalized to volume. Each point corresponds to a common p-stop sensor and the current measured at room temperature (23 °C) was corrected to 20 °C. We obtained a value for the leakage current damage constant $\alpha$ of $(2.08 \pm 0.3) \times 10^{-17}$ A/cm. This is comparable to previous measurements [18]. However, it must be noted that these various measurements were taken under a wide variety of conditions.

We repeated this analysis with the individual p-stop sensors and we found the same behavior. Based on the electrical characterization tests, we do not see any difference between the two p-isolation layouts. We plan to study charge collection in a test beam for both types of sensors before and after irradiation.

Since the irradiated sensors had different guard ring structures (See Table II.), we also checked these for differences in the electrical characteristic after irradiation. No significant dependence was detected for the three guard ring structures.

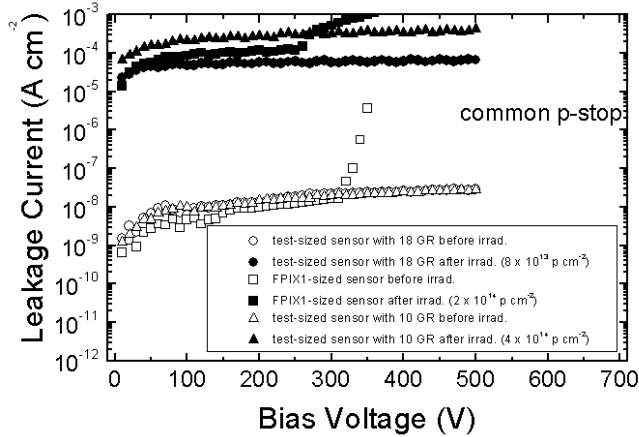

Fig. 5. Leakage current measurements before and after irradiation using non-oxygenated sensors. In this plot, the current is normalized to the sensors' active areas. These measurements were performed at 23°C.

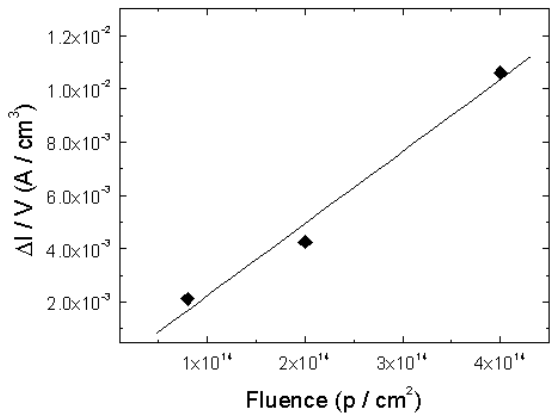

Fig. 6. Fluence dependence of the increase in leakage current. The current was measured at room temperature (23 °C).

As seen in Fig. 5, the current after irradiation increases by a few orders of magnitude. However, operating at lower temperature can alleviate this problem. The measurements shown in Fig. 5 were done at 23 °C. We repeated the same measurements at various temperatures (10 °C, 0 °C and -10 °C) and, as expected, we observed that the current decreases exponentially with temperature ($I_{leak} \cong T^2 \exp(-E / 2k_B T)$). Fig. 7 shows the comparison between data and the predicted dependence of the leakage current vs temperature. There is good agreement between the fit and the data. The current values are for two test-sized common p-stop sensors with 18 guard rings and are measured at the full depletion voltage (~ 80 V for the common p-stop with 18 guard rings and ~130 V for the common p-stop with 10 guard rings). We found that the values of the parameter E are 1.09 eV for the sensor irradiated to $4 \times 10^{14}$ p/cm$^2$ and 1.20 eV for the sensor irradiated to $8 \times 10^{13}$ p/cm$^2$, compatible within the fit errors.

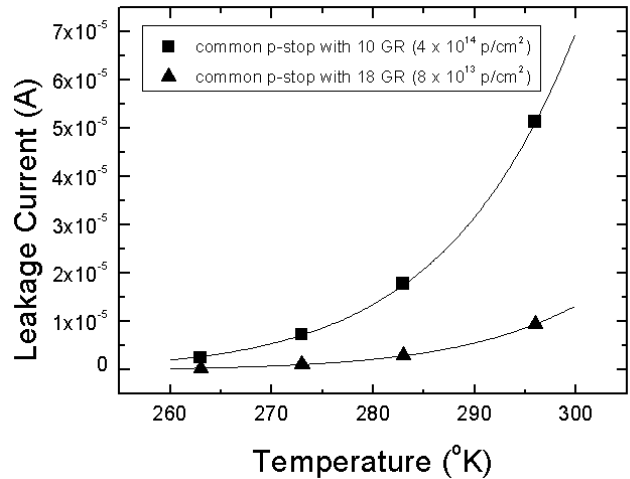

Fig. 7. Leakage current as a function of the temperature for two test-sized common p-stop sensors, one irradiated to $8 \times 10^{13}$ p/cm$^2$ and one to $4 \times 10^{14}$ p/cm$^2$.

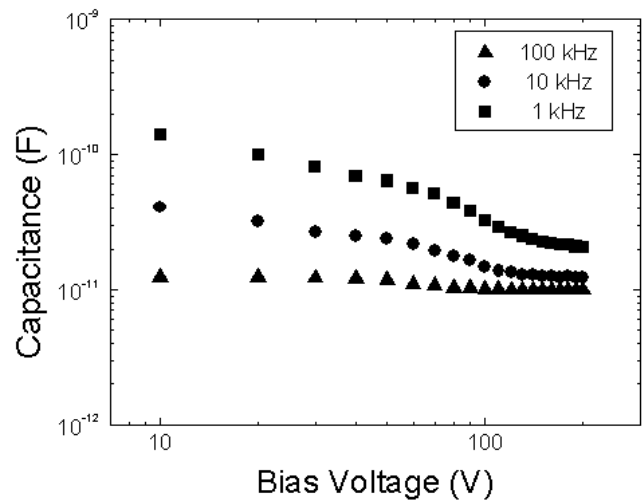

Fig. 8. C-V curves for an irradiated ($4 \times 10^{14}$ p/cm$^2$) test-sized individual p-stop sensor at various frequencies.

The dependences of the depletion voltage and capacitance on the frequency and temperature were also studied. It is well known that after irradiation, CV characteristics have a strong dependence on measurement frequency [19]-[20] and temperature. Preliminary tests were performed on non-irradiated sensors and, as expected, we found no dependence of the C-V characteristics on the measurement frequency used. Fig. 8 shows the C-V measurements at three different frequencies and Fig. 9 shows measurements at four temperatures for an individual p-stop pixel sensor with 10 guard rings after irradiation to $4 \times 10^{14}$ p/cm$^2$. A logarithmic change in frequency gives the same pattern of C-V curves as a linear change in temperature as reported by others [21]-[22].

From Fig. 8 we can see not only the capacitance dependence on the frequency but that the depletion voltage increases as the frequency is decreased. The depletion voltage also decreases with decreasing temperature down to 0 $^{o}$C or so, in agreement with results reported by another group [21].

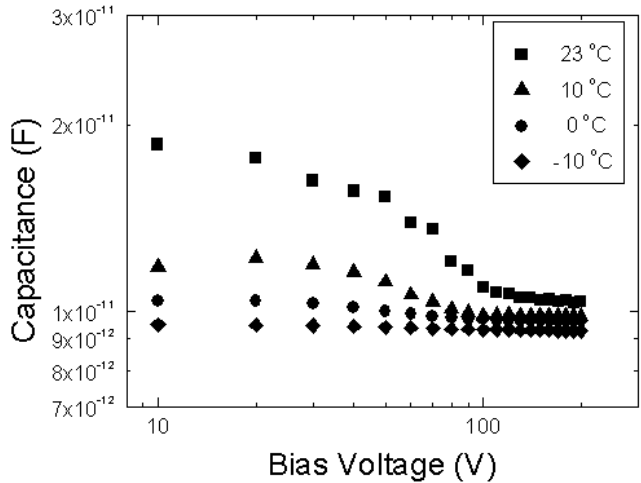

Fig. 9. C-V curves for an irradiated ($4 \times 10^{14}$ p/cm$^2$) test-sized individual p-stop sensor at various temperatures. Measurements were done at a frequency of 40 kHz.

We also studied the voltage distribution over the guard rings. In Fig. 10 we present the results for an oxygenated FPIX1-sized common p-stop sensor (11 guard rings) before and after irradiation. Results indicate that guard rings help to improve breakdown voltage by distributing the potential drop over a longer distance, thus reducing the electric field concentration near the junction boundaries. However, as we can see from the second plot, there is still a potential drop across the device edge after type inversion and more investigation is needed. We repeated the same measurements on a test-sized common p-stop with 18 guard rings irradiated to $4 \times 10^{14}$ p/cm$^2$ and we also find a significant voltage drop.

### C. Oxygenated Sensors

Electrical characterizations were made also for several oxygenated SINTEF wafers. We tested these wafers before and after dicing and some of the single sensors before and after irradiation. For these sensors, as for the standard sensors, the leakage current before irradiation is very small.
We irradiated four oxygenated sensors: two test-sized sensors and two FPIX1-sized sensors (See Table II for details.). Fig. 11 shows the I-V characteristics before and after irradiation (at room temperature) for the common p-stop pixel sensors. We found the same results also for the individual p-stop sensors that we irradiated. As we saw for the non-oxygenated sensors, the test-sized sensors have very good characteristics before and after irradiation (leakage current plateau, high breakdown voltage). The FPIX1-sized sensors on the other hand, have, also in this case, non-optimal performance in breakdown voltage.

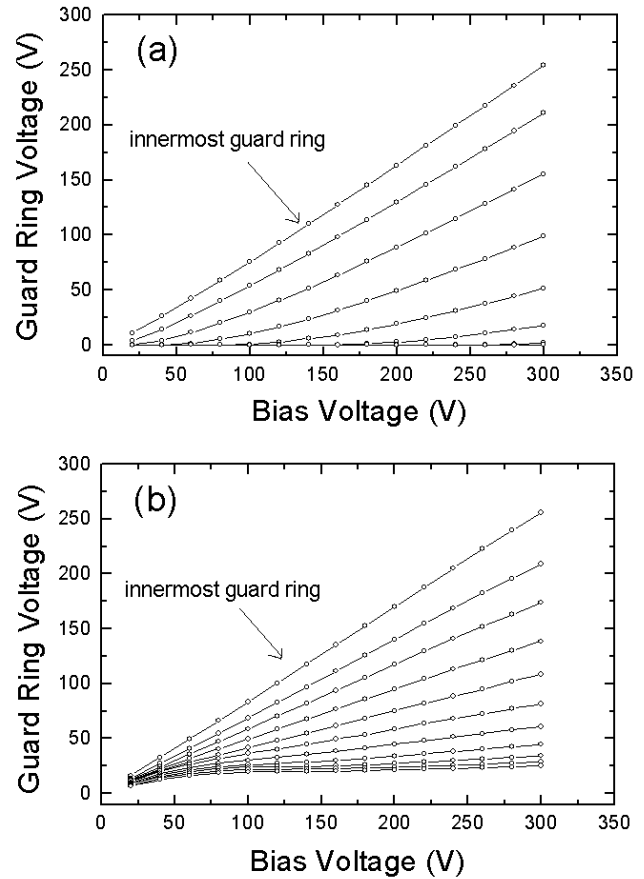

Fig. 10. Potential distribution over the 11 guard rings on the p$^+$ side before a) and after b) irradiation to 2 x $10^{14}$ p/cm$^2$ using an FPIX1-sized common p-stop oxygenated sensor.

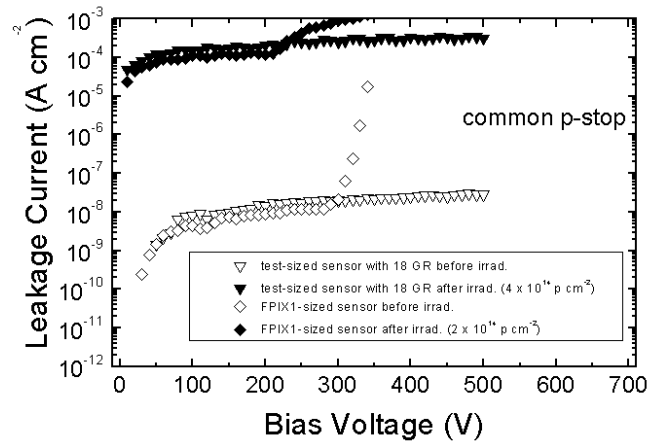

Fig. 11. I-V characteristics (at room temperature) before and after irradiation for oxygenated sensors. In this plot the current is normalized to the sensor's active area.

Fig. 12 shows the dependence of the full depletion voltage on the proton fluences for the normal and the oxygenated sensors. We see that the full depletion voltage at $4 \times 10^{14}$ p/cm$^2$ is still very low, lower than the value before the irradiation. This characteristic is due to the low resistivity of the silicon. This result, together with the fact that the breakdown voltage is still high compared to the full depletion voltage, is very important for the BTeV experiment because we can have fully depleted detectors without biasing at very high voltage. Even though the breakdown voltage for the FPIX1-sized sensors is below 300V after irradiation, it is still well above the depletion voltage, even up to a fluence of $4 \times 10^{14}$ p/cm$^2$.

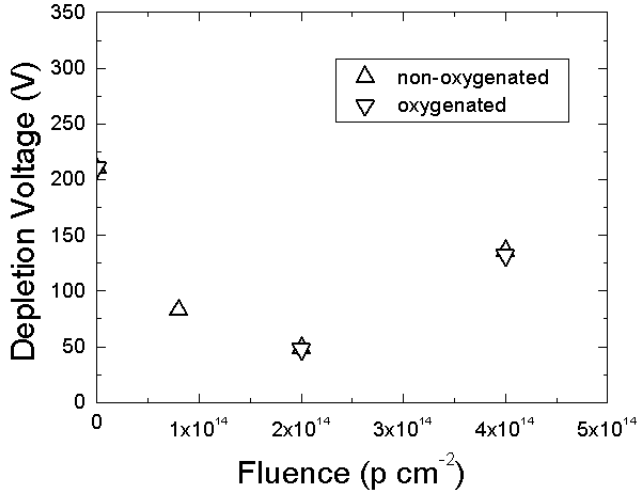

Fig. 12. Depletion voltage as function of proton fluences for normal and oxygenated sensors (270 μm thickness).

We still have some uncertainty on the type inversion point. However from Fig. 12 we can see that, after type inversion, the slope is similar for both oxygenated and non-oxygenated sensors. We can conclude that for these SINTEF sensors, no difference in electrical characteristics before and after irradiation between oxygenated and standard sensors have been observed. We found that the non-oxygenated sensors are as radiation hard as the oxygenated ones. This is in agreement with recent studies performed by other groups, which showed that while there is a large variation in the irradiation results obtained using standard silicon wafers from different foundries, the oxygenation process removes this variation. All the oxygenated wafers show the same performance after irradiation independent of foundry. As here, however, no difference in irradiation results between standard and oxygenated SINTEF diodes was found [23].

## V. CONCLUSION

Experimental results based on I-V and C-V measurements for the prototype BTeV SINTEF pixel sensors are promising. Most of the tested sensors meet the specifications: leakage current less than 50 nA/cm$^2$ and breakdown voltage above 300 V, both for normal and oxygenated sensors before irradiation. Moreover, good results came from a variety of multi-guard ring structures. Very high breakdown voltage protection occurs already with 10 rings. After irradiation, the leakage current significantly increases. However, operating at reduced temperature can minimize the problems associated with the large leakage current. No significant difference was detected between common and individual p-stop isolation. However, the breakdown voltage, both before and after irradiation, appears to depend on the width of the p-implants and/or the gaps between implants. Finally, we detected no difference between the normal and the oxygenated sensors manufactured by SINTEF. The behavior of the full-depletion voltage with the particle fluence is the same for both oxygenated and non-oxygenated sensors, and the values that we measured at $4 \times 10^{14}$ p/cm$^2$ are lower than the values before the irradiation. From the point of view of the radiation hardness with proton fluence, these SINTEF low-resistivity sensors (normal and oxygenated) have excellent performance.

## VI. ACKNOWLEDGMENT

We would like to thank the technical staff of the MicroDetector Group of the Fermilab Technical Centers, in particular to Greg Sellberg, for his support and ingenuity that allowed us to complete this work.